\documentclass[reprint,superscriptaddress]{revtex4-1}
\usepackage{graphicx}
\usepackage{amsmath}
\usepackage{bm}
\usepackage{color}

\usepackage{hyperref}
\hypersetup{%
 pdftitle = {Prediction and near-field observation of skull-guided acoustic waves},
 pdfauthor = {H. Estrada},
 bookmarksopen = true,
 colorlinks  = true,
 citecolor   = black,
 linkcolor   = black,
 raiselinks  = true,
 urlcolor    = blue,
 pdfstartview= FitH,
}

\newcommand{\ksero}{\mathbf{k}_0}
\newcommand{\kserom}{\mathbf{k}_{0m}}

\newcommand{\besselj}[2]{J_{#1}(#2)}

\newcommand{\kvec}[2]{\mathbf{k}_{#1 #2}}

\newcommand{\zu}{\mathbf{\hat{z}}}

\newcommand{\rpar}{\mathbf{r}_\parallel}
\newcommand{\rparu}{\mathbf{\hat{r}}_\parallel}
\newcommand{\kpar}{\mathbf{k}_\parallel}
\newcommand{\rvec}{\mathbf{r}}

\def\ii{{\rm i}}

\begin{document}

\title{Prediction and near-field observation of skull-guided acoustic waves}
\author{H\'ector Estrada}
\affiliation{Institute of Biological and Medical Imaging (IBMI), Helmholtz Center Munich, German Research Center for Environmental Research (GmbH), Ingolstäder Landstra\ss e 1, D-85764 Neuherberg, Germany}
\author{Johannes Rebling}
\affiliation{Institute of Biological and Medical Imaging (IBMI), Helmholtz Center Munich, German Research Center for Environmental Research (GmbH), Ingolstäder Landstra\ss e 1, D-85764 Neuherberg, Germany}
\affiliation{School of Medicine and School of Electrical Engineering and Information Technology, Technical University of Munich, Germany}
\author{Daniel Razansky}
\email{Corresponding author: dr@tum.de}
\affiliation{Institute of Biological and Medical Imaging (IBMI), Helmholtz Center Munich, German Research Center for Environmental Research (GmbH), Ingolstäder Landstra\ss e 1, D-85764 Neuherberg, Germany}
\affiliation{School of Medicine and School of Electrical Engineering and Information Technology, Technical University of Munich, Germany}

\date{\today}
\begin{abstract}
Ultrasound waves propagating in water or soft biological tissue are strongly reflected when encountering the skull, which limits the use of ultrasound-based techniques in transcranial imaging and therapeutic applications. Current knowledge on the acoustic properties of the cranial bone is restricted to far-field observations, leaving its near-field properties unexplored. We report on the existence of skull-guided acoustic waves, which was herein confirmed by near-field measurements of optoacoustically-induced responses in ex-vivo murine skulls immersed in water. Dispersion of the guided waves was found to reasonably agree with the prediction of a multilayered flat plate model. It is generally anticipated that our findings may facilitate and broaden the application of ultrasound-mediated techniques in brain diagnostics and therapy. 
\end{abstract}

\maketitle 
The skull comprises a solid multilayered bony structure which plays a crucial role in protecting the brain from injuries, setting the intracranial pressure balance and determining many other anatomical and functional properties of a living organism. Due to its mechanical properties, ultrasound waves propagating through water or soft biological tissue are strongly reflected when encountering the skull \cite{Fry1978}. Yet, several applications, such as optoacoustic neuroimaging \cite{yw2014}, ultrasound neuromodulation \cite{Naor2016}, focused ultrasound surgery \cite{DanieWJoTU2014}, and blood brain barrier opening for drug delivery \cite{AryalAAMADDR2014}, are able to make efficient use of the ultrasound waves transcranially, mainly because they only require a one-way transmission of the ultrasound wave through the skull. Nonetheless, the spatial resolution and penetration of these transcranial techniques is severely limited due to the skull's presence.

To this end, characterization of the acoustic properties of the cranial bone has been performed either with the aim of focusing ultrasound deep inside the human brain \cite{Clement2002,ClemeHUiM&B2002,ClemeWHTJotASoA2004,MarquPAMMTFPiMaB2009,PichaSHPiMaB2011,pinton2012} or was otherwise aimed at visualizing the mouse brain vasculature \cite{KneipTERSRJoB2016,Estrada2016}. Both cases consider the ultrasound source and/or the target region to be located far away from the skull, i.e. in its far-field. On the contrary, the near-field of an object contains information that is irremediably lost if the object is scrutinized using only the far-field observations. For exam6ple, in near-field scanning optical microscopy  \cite{Betzig1993}, a spatial resolution higher than the far-field diffraction limit is achieved when scanning in the immediate vicinity of the object. Also holographic techniques \cite{Williams1980} can take advantage of the near-field information. 

In the context of acoustic waves in a solid plate, for which no direct electromagnetic wave analogy exists \cite{estrada2009,estrada2012b}, the near-field properties may considerably change if the object effectively comprises a waveguide. The related physical phenomenon, whose manifestation spans from Rayleigh waves triggered by earthquakes \cite{Ben-Menahem1981} down to microchips \cite{Campbell1998}, was first studied by Rayleigh \cite{rayleigh1885} and Lamb \cite{lamb1917} whereas Stoneley \cite{stoneley1924} and Scholte \cite{scholte1942} have further included the effect of solid/solid and solid/fluid interfaces. Currently, guided acoustic waves (GAWs) are employed in non-destructive testing of plates and pipes \cite{Chimenti1991} or material characterization at micro- and nano-scopic scales \cite{hesjedal2010}. GAWs were also characterized in long cylindrical bones aiming at diagnosing osteoporosis \cite{MoilaIToUFaFC2008,TalmaFM2011}, e.g. by determining the thickness of the human radius with multimode GAWs in the axial transmission configuration \cite{ValleBCLMIToUFaFC2016}. The GAWs were also excited optoacoustically in human radius phantoms \cite{MoilaZKKKPMHTUiMB2014}. However, near-field properties of the skull have not yet been explored.

Here we prove the existence of guided acoustic waves in the murine skull bone by means of water-borne laser ultrasound experiments. The experimental findings are supported by numerical simulations performed using a fluid-loaded flat multilayered plate model.

\begin{figure}[b]
 \includegraphics[width=\columnwidth]{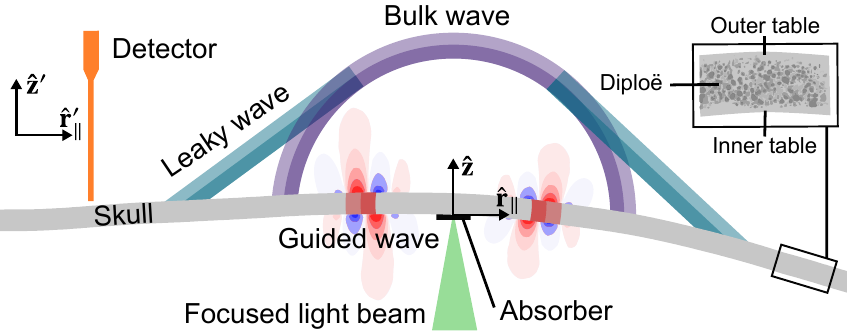}
 \caption{\label{fig:Diagram} Schematic of the skull geometry depicting the acoustic wave propagation mechanisms and the experimental setup. The right inset details the structure of the skull bone. The $(\rparu,\zu)$ coordinates follow the skull curvature, while the $(\rparu^\prime,\zu^\prime)$ do not. The wave propagation in the lower part of the skull has been omitted for clarity.}
\end{figure}

Excitation of a fluid-immersed plate-like structure with a short laser pulse is expected to reveal a number of different wave propagation mechanisms, as depicted in Fig.~\ref{fig:Diagram}. In general, three main mechanisms may exist for ultrasonic wave propagation in the fluid-immersed skull: (a) leaky Lamb-like waves, (b) bulk radiation to the fluid, and (c) guided waves. The first are waves which are radiated into the fluid at an angle given by $sin(\theta_{L})=c_0/c_{L}$ while their propagation along the skull occurs at supersonic speed $c_L>c_0$. 
The second mechanism corresponds to direct radiation from the skull into the surrounding fluid, hence propagating at the speed of sound in the fluid $c_0$. Our underlying hypothesis is that, if the skull further supports a mode having subsonic speed $c_s<c_0$, the corresponding wave will be bound to the skull and will not emit radiation into the far-field. As a result, one may only detect its existence in the near-field.

\begin{figure}[b]
 \includegraphics[width=\columnwidth]{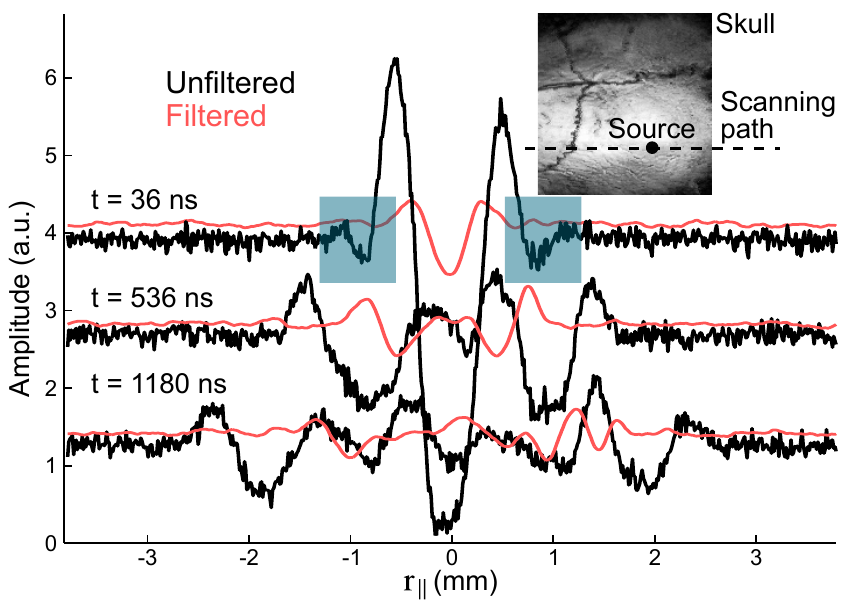}
 \caption{\label{fig:timeSpace} Measured signal amplitude across the scanning path (right inset) for three different time points. Leaky waves are enclosed by blue squares. An animated video with the wave propagation sequence is available \cite{suppMaterial}.}
\end{figure}

Due to obvious experimental challenges involved in direct excitation of ultrasound waves in the skull's near-field, we used a pulsed laser radiation to induce an optoacoustic response in a thin layer of black burnish attached to the interior surface of the skull. The layer effectively acts as a point broadband ultrasound source due to the thermoelastic effect (see Fig.~\ref{fig:Diagram}). A Q-switched diode laser of 532 nm in wavelength (EdgeWave GmbH, W\"urselen, Germany) was used to generate 3 $\mu$J pulses of 10 ns duration, which were focused onto the absorbing film attaining a beam diameter of approximately 100 $\mu$m at the surface. Accurate mapping of the near-field acoustic field is achieved by scanning a 0.5 mm diameter polivynil difluoride (PVdF) needle hydrophone (Precision Acoustics, UK) in close proximity ($<$100 $\mu$m) to the skull surface. Due to the curvature of the mouse skull, the exact three-dimensional scanning pattern  was constructed using high resolution pulse-echo images of the skull surface acquired with a 30 MHz spherically focused PVdF ultrasound transducer (Precision Acoustics, UK), as shown in the inset of Fig.~\ref{fig:timeSpace}. During the experiments, the skull bone of a 6 weeks old mouse was immersed in phosphate buffered saline (PBS) solution (Life Technologies Corp., UK).

The detected ultrasonic signals covering the left frontal, parietal, and occipital bones are depicted in Fig.~\ref{fig:timeSpace} for the different time instants following the laser pulse. The supersonic waves can only be observed shortly after the main bang (36 ns, blue rectangles). 

\begin{figure}[t]
 \includegraphics[width=\columnwidth]{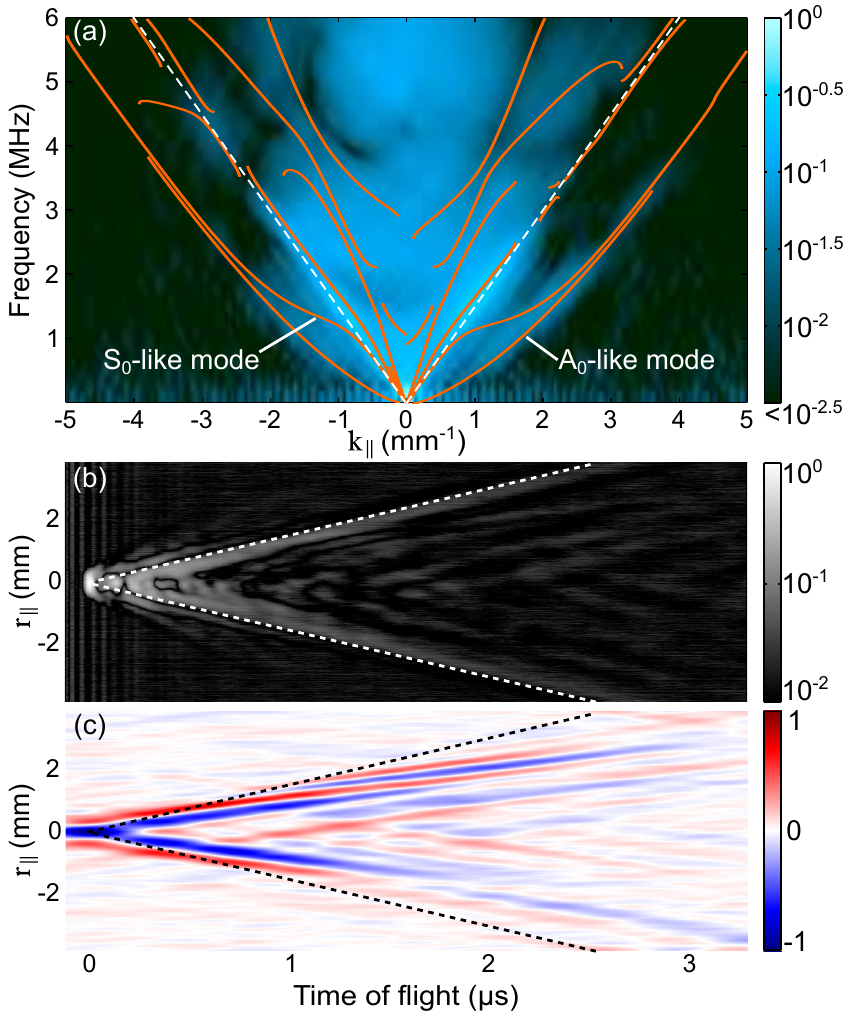}
 \caption{\label{fig:gResults}  (a) Measured and calculated (overlay) dispersion. Raw (b) and filtered (c) wave propagation as a function time and distance. The dashed lines depict the speed of sound in water.}
\end{figure}

The main features of both wavefronts traveling away from the source follow the speed of sound in the fluid $c_0=1502$ m$/$s \footnote{It has been calculated considering the temperature (23 $^o$C) and the salinity (10 g$/$L) of PBS according to \cite{WongZTJotASoA1995}}. Note that the traces of the waves propagating at a subsonic speed are only evident after the spatio-temporal data $(\rpar,t)$ is filtered in reciprocal space $(\kpar,\omega)$. Near- and far-field are then separated by the dashed line representing $c_0$ in Fig.~\ref{fig:gResults}(a) after applying a two-dimensional Fourier transform. The existence of subsonic modes is evident as well the asymmetry regarding the propagation direction, the latter expected due to the inhomogeneous conformation of the skull. 

As a first approximation, we calculated the modal dispersion of a flat viscoelastic plate consisting of three isotropic layers immersed in water. Assuming plane wave propagation and applying appropriate boundary conditions results in a linear system of equations yielding the mode dispersion as a function of $(\kpar,\omega)$ \cite{LoweUFaFCITo1995} (for more details see Appendix A). Overlaid in Fig.~\ref{fig:gResults}(a) is the mode dispersion obtained using the density $\rho$ and longitudinal speed of sound $c_{\ell}$ of the cortical and trabecular bone layers \cite{cgts2010}, while the transverse speed of sound $c_t$ and the volumetric $\zeta$ and shear $\eta$ viscoelastic losses were adjusted to fit the experimental results \footnote{Cortical bone: $\rho=$ 1969 kg$/$m$^3$, $c_{\ell}=$ 3476 m$/$s, $\zeta=$ 1$\times$10$^{-2}$ Pa$\,$s, $c_{t}=$ 1760 m$/$s, $\eta=$ 1$\times$10$^{-1}$ Pa$\,$s. Trabecular bone: $\rho=$ 1055 kg$/$m$^3$, $c_{\ell}=$ 1886 m$/$s, $\zeta=$ 1.5 Pa$\,$s, $c_{t}^+=$ 650 m$/$s, $c_{t}^-=$ 800 m$/$s, $\eta=$ 3 Pa$\,$s}. 
The relative thickness of the individual layers (see inset in Fig.~\ref{fig:Diagram}) was assumed to follow the proportion of 0.26, 0.5, and 0.24 \cite{pinton2012} for the total skull thickness of $h^{-}=$ 320 $\mu$m and $h^{+}=$ 300 $\mu$m in the negative and positive propagation directions, respectively.

Two subsonic Lamb-Rayleigh-like modes nearly overlap for frequencies above 2 MHz and match the experimental results with $800<c_s<1000$ m$/$s. At lower frequencies the modes branch, thus resembling the behavior of cut-off-free symmetric S$_0$ and antisymmetric A$_0$ Lamb modes. By defining a window in the subsonic region as $\omega<0.7c_0\kpar$, using a threshold to reject noise, and applying an inverse two-dimensional Fourier transform, one obtains the filtered spatio-temporal propagation of the subsonic mode (Figs.~\ref{fig:gResults}(c) and \ref{fig:timeSpace}). This type of processing makes the subsonic wave clearly distinguishable, in contrast to the unfiltered data (Fig.~\ref{fig:gResults}(b)).

\begin{figure}[t]
 \includegraphics[width=\columnwidth]{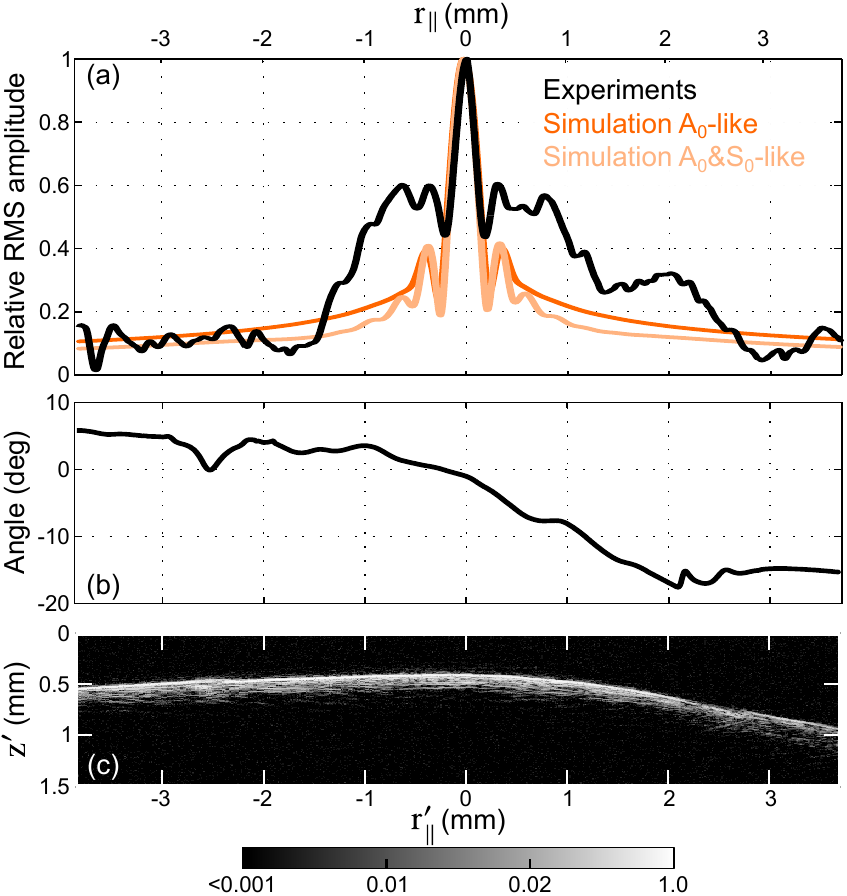}
 \caption{\label{fig:xDecay} (a) Guided wave decay along the propagation direction $\mathrm{r}_\parallel$ extracted from Fig.~\ref{fig:gResults}(c) and compared against the calculations. (b) Angle of the skull surface relative to the scanning direction $\mathrm{r}_\parallel^\prime$. (c) Ultrasound pulse-echo image of the mouse skull in the scanned region.}
\end{figure}

The decay of the guided-wave in the propagation direction $\rpar$ (Fig.~\ref{fig:xDecay}(a)) is subsequently calculated by taking the root-mean-square (RMS) of the filtered wave using an appropriate spatio-temporal window to isolate the propagating wave component from noise and any scattering events. If the effects of the skull inhomogeneities were negligible, one could expect a decay $\propto \text{exp}(-\alpha \rpar)/\sqrt{\rpar}$ \cite{morse_ingard} far away from the source with $\alpha$ being an attenuation constant. In addition, due to the generally broad-band nature of the optoacoustic source, the dispersion (see Fig.~\ref{fig:gResults}(a)) needs to be considered for each individual mode $m$. The pressure in the fluid on top of the skull ($z>h$) can be calculated as
\begin{equation}
 p_m(\rvec,t) = \frac{\ii\rho_0}{4\pi^2}\,\iint_{-\infty}^\infty \Gamma(\kpar)\,\Phi_m(\kpar,z)\,e^{\ii \kpar\cdot\rpar}\,d^2\kpar\,, \label{eq:timeFull}
\end{equation}
where $\Phi_m(\kpar,z) = \omega_m(\kpar)\,\mathrm{exp}\left(\ii k_{mz}( z-h) - \ii\omega_m(\kpar) t\right)$, $\Gamma(\kpar)$ depends on the initial pressure and its time derivative, $k_0 = \omega/c_0+ \ii \alpha_0\omega^2 $ with $\alpha_0/(2\pi) = 22\times10^{-15}$ s$^2$/m \cite{acmic2008}, and $k_{mz}=\sqrt{\kserom^2-\kpar^2}$ (see Appendix A).

The symmetry of the experimental curve is broken by a local maximum around $\mathrm{r}_\parallel = 2$ mm away from the optoacoustic source. The prediction of the flat-plate model is not able to explain the experimentally observed slow decay in the vicinity of the source. The deviation may be attributed to the non-uniform spatial sensitivity of the 0.5 mm diameter detector used for the measurements as well as the acoustic scattering effects at the suture boundaries and other inhomogeneities of the skull. Indeed, the guided wave propagating parallel to the skull surface will reach the detector at different angles due to the skull curvature. The sutures are located at $\mathrm{r}_\parallel=-2.5$ mm and $\mathrm{r}_\parallel=2$ mm. While the suture to the right is involved in the local maximum of the experimental decay curve, the amplitude of the wave propagating to the left approaches noise levels just before reaching the left suture. However, one may also observe in Fig.~\ref{fig:gResults}(c) that the latter wave component continues propagating also for $\mathrm{r}_\parallel<-2$ mm after abruptly changing its amplitude. 

\begin{figure}[h!]
 \includegraphics[width=\columnwidth]{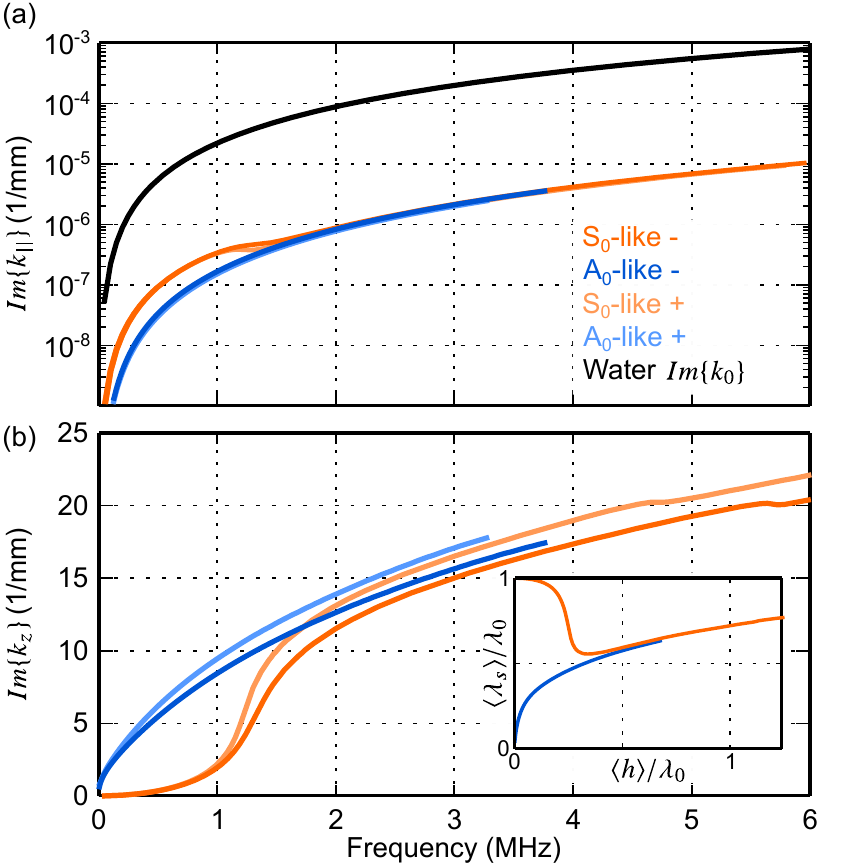}
 \caption{\label{fig:attenuation} Imaginary part of the wavenumber in the (a) $\rparu$ and the (b) $\zu$ directions as a function of the frequency. The attenuation in water is plotted in (a) as reference. The inset in (b) shows the ratio between the subsonic wavelength averaged for the $+$ and $-$ direction $\langle \lambda_s \rangle$ and the wavelength in the fluid $\lambda_0$ as a function of the averaged skull thickness $\langle h \rangle$ normalized as well by $\lambda_0$.}
\end{figure}

Furthermore, close inspection of the imaginary part of the wavenumber $Im\{k_{\parallel}\}$ reveals that the losses due to viscosity play a relatively minor role in the propagation of the skull-guided-waves within the studied frequency range (see Fig.~\ref{fig:attenuation}(a)) since the effective losses are orders of magnitude weaker as compared to the corresponding losses in water for bulk wave propagation. This made it difficult the selection of appropriate viscosity constants as both negligible and high volumetric $\zeta$ and shear viscosities $\eta$ effectively gave rise to the same propagation scenario. 
The attenuation in the $\zu$ direction remains also nearly unaffected by changes in the viscoelastic losses, as it is mostly determined by the geometrical relation between $\ksero$ and $\kpar$. Figure \ref{fig:attenuation} shows the different behavior for S$_0$ and A$_0$-like modes, particularly at the low frequency range. As a reference, 10 mm$^{-1}$ results in a characteristic half-decay distance of about 70 $\mu$m.      

Although our simplified model generally matches the experimental dispersion, limitations in the experimental spatio-temporal resolution along with additional effects induced by skull’s curvature and inhomogeneities have made some of the measured modes difficult to discern. Particularly challenging is the accurate distinction between the mode located close to the sound line and the bulk wave mode propagating at a grazing angle (see Fig.~\ref{fig:gResults}(a)). 

In general, the guided wave has been observed here to propagate for distances beyond 3 mm across the mouse skull. Although the experimental measurements are affected by scattering and artifacts due to the finite size of the detector and the skull curvature, the flat-plate model has been found to agree within an order of magnitude with the experimental results. In addition, the theoretical model clearly points towards curvature and scattering as the main sources of attenuation for the skull-guided wave propagation. 

From a practical stand-point, the reported skull-guided waves could be potentially used as sole or complementary carriers of acoustic information across the skull in imaging or therapeutic applications. Due to similitudes of the cranial bone structure in small mammals and humans \cite{KneipTERSRJoB2016}, it is generally expected that similar phenomena also exists in human skulls, although must be scaled accordingly in frequency and space. Acoustic scattering and exponential decay along the depth direction ($\zu$) may restrict the practical use of the guided-wave phenomenon to low frequencies. However, the achievable imaging resolution would not be compromised due to the shorter wavelength of the subsonic waves in comparison to bulk longitudinal waves in water for the same frequency (see inset in Fig.~\ref{fig:attenuation}(b)). Although the dispersive nature of the skull-guided waves has to be further considered, the propagation problem is effectively 2.5-dimensional. Naturally, penetration depth is limited by the evanescent decay in the $\zu$ direction, yet could be sufficient to target cortical brain structures per-cranially.

Parameters of the immersion fluid are crucial in determining whether the particular mode remains guided or turns leaky. In this regard, our choice for water as a coupling medium is not random as its acoustic impedance is very close to most soft tissues \cite{acmic2008} including the brain, while water-based gels are further used as coupling medium in many biological applications. However, if the skull is surrounded on its either side by a different medium, e.g. air, the wave phenomena may dramatically differ.   

In summary, we report for the first time, to the best of our knowledge, on the existence of guided acoustic waves in skull bone surrounded by water-like media. Solid experimental evidence was further obtained on the dispersion of the skull-guided waves by means of near-field mapping of optoacoustically-generated broadband waves. Reasonable agreement was found with the predictions of a multilayered flat plate model. Characteristic decays in both the lateral and depth directions were characterized both experimentally and theoretically. Due to the general scalability of our model and the experimental observations, the reported guided waves are also expected to exist in human skulls. Thus, our findings can be used to facilitate and broaden the application of ultrasound-mediated techniques in brain diagnostics and therapy. 

\acknowledgements{The research leading to these results has received funding from the European Research Council under grant agreement ERC-2015-CoG-682379. D. R. and J. R. acknowledge support from the European Research Council through the Initial Training Network Grant Nr. 317526.}

\appendix
\section{Flat multilayered viscoelastic plate model}
We used the global matrix method \cite{LoweUFaFCITo1995} to simulate the skull behavior to first approximation. We first divide the space into three flat isotropic solid layers of thickness $h=\sum_{j=1}^3 h_j=\sum_{j=1}^3(z_{j}-z_{j-1})$ sandwiched by two fluid semi-infinite spaces. Then, as a solution to the Helmholtz equation, we assume plane (inhomogeneous) waves in each layer with $\mathrm{ exp}(-\ii\omega t)$ time dependence. For $v=\{\ell,t\}$ corresponding to longitudinal and transverse waves, respectively, the complex bulk wavenumbers $k_{vm}=\omega/c_{vm}+\ii\alpha_{vm}(\omega)$ determine the wavevectors $\mathbf{k}_{vm}^{\pm}=k_{vm}\left( \sin(\theta_{vm})\rparu\pm \cos(\theta_{vm})\zu\right)$, with the polar angle $\theta_{vm}$. For simplicity, the fluids 0 and 4 are considered the same and the subindex $v$ is omitted as only longitudinal waves can propagate on it.

The plane (inhomogeneous) wave potentials $\phi_0\,,\:\phi_4$ represent the wave propagation in the fluid, whereas in the solid layers $j=1,2,3$, $\psi_{vj}$ describes both longitudinal and transverse plane (inhomogeneous) wave potentials. Setting $z=0$ at the first fluid-solid interface with $\zu$ pointing upwards (Fig.~\ref{fig:Diagram}) we can explicitly write  
\begin{eqnarray}
  \phi_0&=&Ie^{i\kvec{}{0}^{+}\cdot\rvec}+Re^{i\kvec{}{0}^{-}\cdot\rvec}\,,\label{eq:presdefhomo1}\\
  \psi_{v j}&=&A_{vj}^{+}e^{i\kvec{v}{j}^{+}\cdot(\rvec-\mathbf{z}_{j-1})}+A_{vj}^{-}e^{i\kvec{v}{j}^{-}\cdot(\rvec-\mathbf{z}_j)}\,,\label{eq:presdefhomo:2}\\
  \phi_4&= &Te^{i\kvec{}{0}^{+}\cdot(\rvec-\mathbf{h})}\,,\label{eq:presdefhomo3}
\end{eqnarray} 
where $I,\: R,\: T$ correspond to the incident, reflected, and transmitted complex wave amplitudes, respectively. In the solid, volumetric $\zeta$ and shear $\eta$ viscosities are included in the Lam\'e constants as $\lambda=\lambda_0+\ii \omega(2\eta/3-\zeta)$ and $\mu=\mu_0-\ii\omega\eta$ to account for absorption \cite{brekhovskikhALM}, which is reflected in the imaginary part of the bulk wavenumbers. 
Applying continuity of the displacement and the stresses at each boundary we obtain a system of 14 equations were the unknowns are $R(\kpar,\omega),\:T(\kpar,\omega),\:$and $ A_{vj}^{\pm}(\kpar,\omega)$. 
The calculation starts by solving the transmission problem for $I=1$ and real $k_{\parallel}$ in a given $(k_{\parallel},\,\omega)$ range and step size $(\Delta k_{\parallel},\,\Delta\omega)$. Once the transmission maxima are identified with the different modes $m$, the $(k_{\parallel},\,\omega_m^0(k_{\parallel}) \pm\Delta\omega)$ values serve as an input to a golden section search for the $(k_{\parallel},\,\omega_m(k_{\parallel}))$ pair (now with complex $k_{\parallel}$) that produces a singular matrix under the condition $I=0$. A thoroughly explanation on a similar method can be found in \cite{Lowe1992}. 

Once the mode's dispersion is known, the pressure in the fluid on top of the skull can be expanded as $p(\rvec,t) = \sum_m p_m(\rvec,t)$. Following \cite{jackson1999}, we can calculate the fate of an initial pressure field $p(0)\equiv p(\rpar,h,0)$ and its derivative $\partial_t p(0)\equiv\partial_t p(\rpar,h,t)|_{t=0}$ as

\begin{equation}
 p_m(\rvec,t) = \frac{\ii\rho_0}{4\pi^2}\,\iint_{-\infty}^\infty \Gamma(\kpar)\,\Phi_m(\kpar,z)\,e^{\ii \kpar\cdot\rpar}\,d^2\kpar\,, \label{eq:timeFull}
\end{equation}
where $\Gamma(\kpar)$ depends on the initial conditions, 

\begin{equation}
 \Phi_m(\kpar,z) = \omega_m(\kpar)\,e^{\ii (k_{mz}( z-h) - \omega_m(\kpar) t)}\,,
\end{equation}

and $k_{mz}=\sqrt{\kserom^2-\kpar^2}$. As the solid is assumed to be isotropic, one can take advantage of the symmetry around the azimuthal angle $\varphi$ and obtain
\begin{equation}
 p_m(\rvec,t) = \frac{\ii\rho_0}{2\pi}\,\int_0^\infty \Gamma(k_r)\,\Phi_m(k_r,z)\besselj{0}{k_r}k_r\,dk_r\,,  \label{eq:timeHankel}
\end{equation}
 which corresponds to a zero-order Hankel transform over the wavenumber in the radial direction $k_r$. Using the shorthand notation from \cite{fourieracoustics} it becomes
\begin{equation}
 p_m(\rvec,t) = \frac{\ii\rho_0}{2\pi}\,\mathcal{B}_0^{-1}\left\{ \Gamma(k_r)\,\Phi_m(k_r,z)\right\} \,.  \label{eq:timeHankelshort}
\end{equation}
Thus, we can write 
\begin{eqnarray}
 \Gamma(k_r) =  \frac{2\pi}{\ii\rho_0}\,\left(\frac{\mathcal{B}_0\left\{\partial_t p(0)\right\}}{\omega^2(k_r)} 
-\frac{\ii\mathcal{B}_0\left\{p(0)\right\}}{\omega(k_r)} \right)\,. \label{eq:initialSpectrum}
\end{eqnarray}

\begin{figure}[h!]
 \includegraphics[width=\columnwidth]{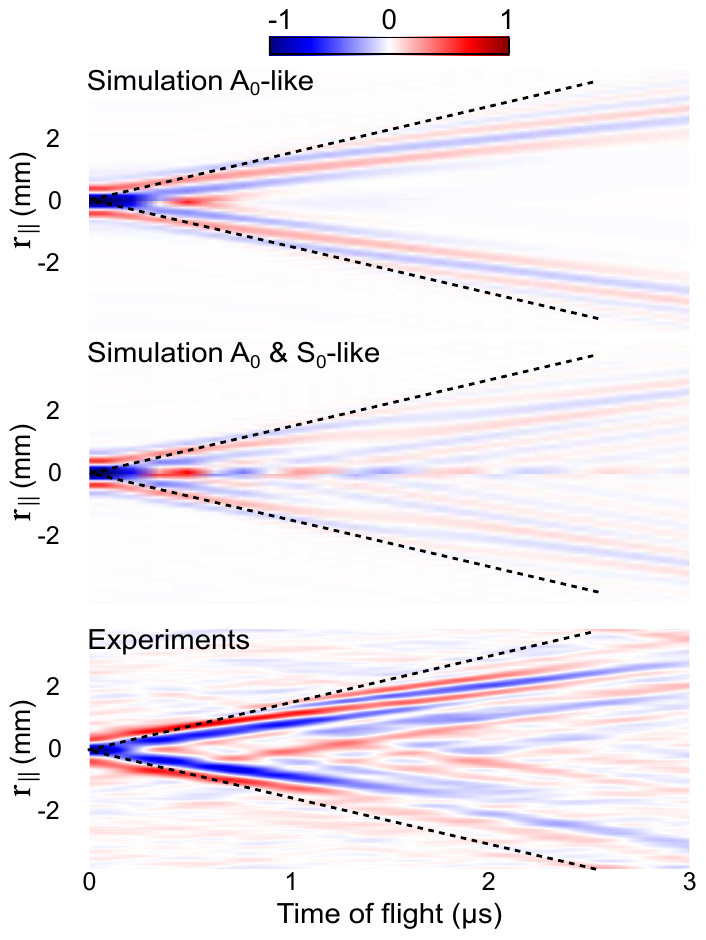}
 \caption{\label{fig:simulatedField} Spatio-temporal skull-guided-wave propagation from simulations ((a), (b)) and experiments (c). The color scale represents the normalized pressure and the dashed black line the speed of sound in the fluid.}
\end{figure}

The numerical evaluation of Eq.~\ref{eq:timeHankel} is performed using Bessel series expansion \cite{GuizarSicairos2004} with 1500 points in total, distributed in 15 mm to avoid unwanted reflections that arise when this method is used. $\Gamma(k_r)$ was extracted from the filtered experimental data from Fig.~\ref{fig:gResults}(c). The results of the simulations using the different viscoelastic constants for the negative and positive propagation direction are shown in Fig.~\ref{fig:simulatedField}. Calculation of the RMS pressure across time for Fig.~\ref{fig:simulatedField}(a) and (b) yields the theoretical curves from Fig.~\ref{fig:xDecay}(a). 

\section{Spatial sensitivity of a circular detector}
\begin{figure}[h!]
 \includegraphics[width=\columnwidth]{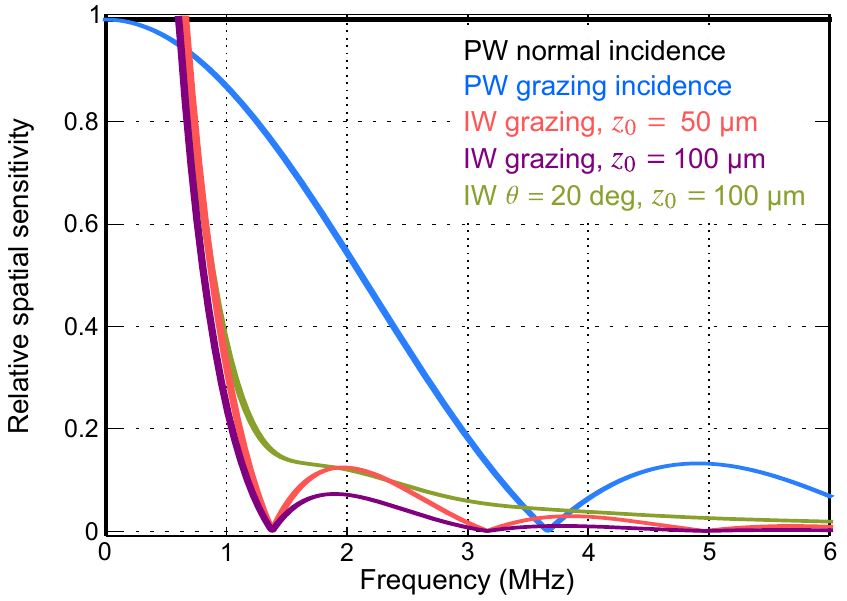}
 \caption{\label{fig:sensitivity} Relative spatial sensitivity as a function of frequency for plane waves (PW) and inhomogeneous plane waves (IW) as described by the labels.}
\end{figure}

We can calculate the effect of the detector's finite size and geometry on near-field experiments, particularly when measuring evanescent waves. For a circular detector of radius $R$ detecting plane waves of amplitude $\phi_0$ and wavevector $\ksero$, the sensitivity $\Pi_{\mathrm{PW}}(k_r)$ is given by

\begin{eqnarray}
 &\Pi_{\mathrm{PW}}(k_r) = &\phi_0\int_0^{2\pi}\int_0^R e^{\ii\ksero\cdot\rvec} r\,dr\,d\varphi\,,\nonumber\\
 &\Pi_{\mathrm{PW}} (k_r) = &\phi_0\pi R^2 \begin{cases}
                    1& \text{if }k_r=0\,,\\
		  \frac{2\besselj{1}{k_r R}}{ k_r R}& \text{if }k_r\neq 0\,.
                  \end{cases}
 \label{eq:planeWaveSens}
\end{eqnarray}
The well known result for normal incidence ($k_r=0$) with the detector's sensitivity being proportional to its area is recovered.
Now, for an inhomogeneous plane wave propagating on top of a flat surface placed at a distance $z_0$ and rotated an angle $\theta$ relative to the detector's normal vector

\begin{eqnarray}
 &\Pi_{\mathrm{IW}}(z_0,\theta,\omega) = &\phi_0\int_0^{2\pi}\int_0^R e^{(i\kpar-\mathbf{k}_z)\cdot\rvec^\prime} r^\prime\,dr^\prime\,d\varphi^\prime\,, \nonumber\\
 &\Pi_{\mathrm{IW}}(z_0,\theta,\omega)  = &\phi_0\pi R^2\, e^{\ii q_r^* z_0} \frac{2\besselj{1}{q_r R}}{ q_r R} \,,                  
 \label{eq:inhomoWaveSens}
\end{eqnarray}

where $q_r = k_{\parallel}\cos{\theta}-\ii k_z\sin{\theta}$, $q_r^* = k_{\parallel}\sin{\theta}-\ii k_z\cos{\theta}$, and $k_{z}=\sqrt{\ksero^2-\kpar^2}$. Taking $k_{\parallel} = \omega/c_s$ with $c_s = $ 900 m/s, $R = $ 0.5 mm and normalizing by $\phi_0\pi R^2$ we obtain the relative spatial sensitivity for different cases shown in Fig.~\ref{fig:sensitivity}. The detection of inhomogeneous plane waves seems more efficient than for plane waves in the low frequency regime. However, there is a strong cut-off below 1.5 MHz for inhomogeneous plane waves at different $z_0$ and $\theta$. This simple calculation is valid only far away from the source and serves to illustrate how the detector affects the measurements in the skull's near-field. Close to the source, the exact shape of the field has to be taken in account. 

\end{document}